\documentclass[structabstract]{aa}  
\pdfoutput=1
\usepackage{graphicx}
\usepackage{epstopdf}
\usepackage{ulem}
\usepackage{txfonts}
\usepackage{natbib}
\begin{document}

   \title{CNO abundances and carbon isotope ratios in evolved stars of the open clusters NGC\,2324, NGC\,2477, and  NGC\,3960\thanks{Based on observations collected at ESO telescopes under programmes 072.D-0550 and 074.D-0571.
}
}

   \author{
 Gra\v{z}ina Tautvai\v{s}ien\.{e}\inst{1},      
Arnas Drazdauskas\inst{1},
         Angela Bragaglia\inst{2},
      Sofia Randich\inst{3},
          \and
     Renata \v{Z}enovien\.{e}\inst{1}                
 }
   \institute{Institute of Theoretical Physics and Astronomy, Vilnius University,
              A. Gostauto 12, LT-01108 Vilnius\\
              \email{grazina.tautvaisiene@tfai.vu.lt }
         \and
            INAF - Osservatorio Astronomico di Bologna, Via Ranzani 1, I-40127 Bologna, Italy
         \and
         INAF - Osservatorio Astrofisico di Arcetri - Largo Enrico Fermi 5, I-50125 Firenze, Italy
             }

\authorrunning{G. Tautvai\v{s}ien\.{e} et al.}
\titlerunning {Open clusters NGC\,2324, NGC\,2477, and  NGC\,3960}

 
  \abstract
   {}
   {Our main aim is to determine carbon-to-nitrogen and  carbon isotope ratios for evolved giants in the open clusters NGC\,2324, NGC\,2477, and  NGC\,3960, which have turn-off masses of about 2~$M_{\odot}$, and to compare them with predictions of theoretical models.}
   {High-resolution spectra were analysed using a differential synthetic spectrum method. Abundances of carbon were derived using the ${\rm C}_2$ Swan (0,1) band heads at 5135 and 5635.5~{\AA}. The wavelength interval 7940--8130~{\AA} with strong CN features was analysed to determine nitrogen abundances and carbon isotope ratios. The oxygen abundances were determined from the [O\,{\sc i}] line at 6300~{\AA}.}
   {The mean values of the CNO abundances are ${\rm [C/Fe]}=-0.35\pm0.06$ (s.d.), ${\rm [N/Fe]}=0.28\pm0.05$, and ${\rm [O/Fe]}=-0.02\pm0.10$ in seven stars of  NGC\,2324;  ${\rm [C/Fe]}=-0.26\pm0.02$, ${\rm [N/Fe]}=0.39\pm0.04$, and  ${\rm [O/Fe]}=-0.11\pm0.06$ in six stars of NGC\,2477; and ${\rm [C/Fe]}=-0.39\pm0.04$, ${\rm [N/Fe]}=0.32\pm0.05$,
and  ${\rm [O/Fe]}=-0.19\pm0.06$ in six stars of NGC\,3960. The mean C/N ratio is equal to $0.92\pm 0.12$, $0.91\pm 0.09$, and $0.80\pm 0.13$, respectively. The mean $^{12}{\rm C}/^{13}{\rm C}$ ratio is equal to $21\pm 1$, $20\pm 1$, and $16\pm 4$, respectively. The $^{12}{\rm C}/^{13}{\rm C}$ and C/N ratios of stars in the investigated open clusters were compared with the ratios predicted by stellar evolution models.}
  {The mean values of the $^{12}{\rm C}/^{13}{\rm C}$ and C/N ratios in NGC\,2324 and NGC\,2477 agree well with the first dredge-up and thermohaline-induced extra-mixing models, which are similar for intermediate turn-off mass stars. The $^{12}{\rm C}/^{13}{\rm C}$ ratios in the investigated clump stars of NGC\,3960 span from 10 to 20. The mean carbon isotope and C/N ratios in NGC\,3960 are close to predictions of the model in which the thermohaline- and rotation-induced (if rotation velocity at the zero-age main
sequence was 30\% of the critical velocity) extra-mixing act together.}

   \keywords{stars: abundances --
                stars: horizontal branch --
                stars: evolution --
                open clusters and associations: individual:  NGC\,2324, NGC\,2477, and  NGC\,3960
               }

   \maketitle

\section{Introduction}

It is recognised that the first dredge-up (1DUP) is not the only mixing event during the red giant branch (RGB) ascent that alters the surface composition of a star (\citealt{Chaname05}, and references therein). There should be some form of extra-mixing that brings products of partial hydrogen burning into the stellar envelope. Many possible physical mechanisms might be involved: rotational mixing (\citealt{Sweigart79, Palacios03, Chaname05, Denissenkov06}), magnetic fields (\citealt{Hubbard80, Busso07, Nordhaus08, Palmerini09}), rotation and magnetic fields (\citealt{Eggengerger05}), internal gravity waves (\citealt{Zahn97, Denissenkov00}), thermohaline mixing (\citealt{Eggleton06, Eggleton08, Charbonnel07, Cantiello10, Charbonnel10}), combination of thermohaline mixing and magnetic fields \citep{Busso07, Denissenkov09}, and a combination of thermohaline mixing and rotation (\citealt{Charbonnel10, Lagarde12}). 

The thermohaline-induced mixing is not optional (cf. \citealt{Eggleton08}); it inevitably arises on the first-ascent giant branch, when the hydrogen-burning shell encroaches on the homogenized, formerly convective zone left behind by the retreating convective envelope, and begins to burn $^3$He. This $^3$He burning occurs just outside the normal hydrogen-burning shell, at the base of a radiatively stable region of about 1~$R_{\odot}$ in thickness. The burning causes a molecular weight inversion, which drives mixing all the way to the convection zone. It not only destroys about 90\% of the $^3$He produced in the low-mass stars, but reduces the $^{12}{\rm C}/^{13}{\rm C}$ and C/N ratios. The rotation-induced mixing modifies the internal chemical structure of main-sequence stars, although its signatures are revealed only later in the evolution when the first dredge-up occurs. It favours the occurrence of extra mixing in RGB stars in the mass range between about 1.5 and 2.2~$M_{\odot}$ (cf. \citealt{Charbonnel10}). 

In this work, we aim to determine C N O abundances and $^{12}{\rm C}/^{13}{\rm C}$ ratios for red giants in three open clusters: NGC\,2324, NGC\,2477, and  NGC\,3960. The turn-off masses of these clusters are about 2~$M_{\odot}$ (\citealt{Sestito06, Bragaglia08}). This mass range has previously not been well covered, therefore these clusters are suitable targets in which to study the dispersion of abundances that might be caused by rotation-induced mixing in stars of similar mass and evolutionary status. We analyse  part of the same spectra as used by \citet{Sestito06} and \citet{Bragaglia08} in their studies.

Investigations of NGC\,2324 have begun with \citet{Cuffey41}, who built a colour-magnitude diagram and estimated a distance of about 3320 parsecs. NGC\,2324 is a relatively young cluster. \citet{Kyeong01} determined an age of about 630~Myr, a Galactocentric distance $R_{\rm gc}=11.7$~kpc, and [Fe/H]$\sim -0.32$ using $UBVIc$ CCD photometry and isochrone fitting. More recently, \citet{Piatti04} used $VIc$ and $Washington$ photometry to derive an age of about 440~Myr, a distance of 3.8 kpc, reddening 0.25, and metallicity of about $-0.30$~dex.  The analysis of low-resolution spectra reported by \citet{Friel02} provided [Fe/H]=$-0.15 \pm 0.16$, while the analysis of high-resolution spectra, which we use here, has been performed by \citet{Bragaglia08} and produced [Fe/H]=$-0.17 \pm 0.05$. 

  
\begin{figure}
\includegraphics[width=0.47\textwidth]{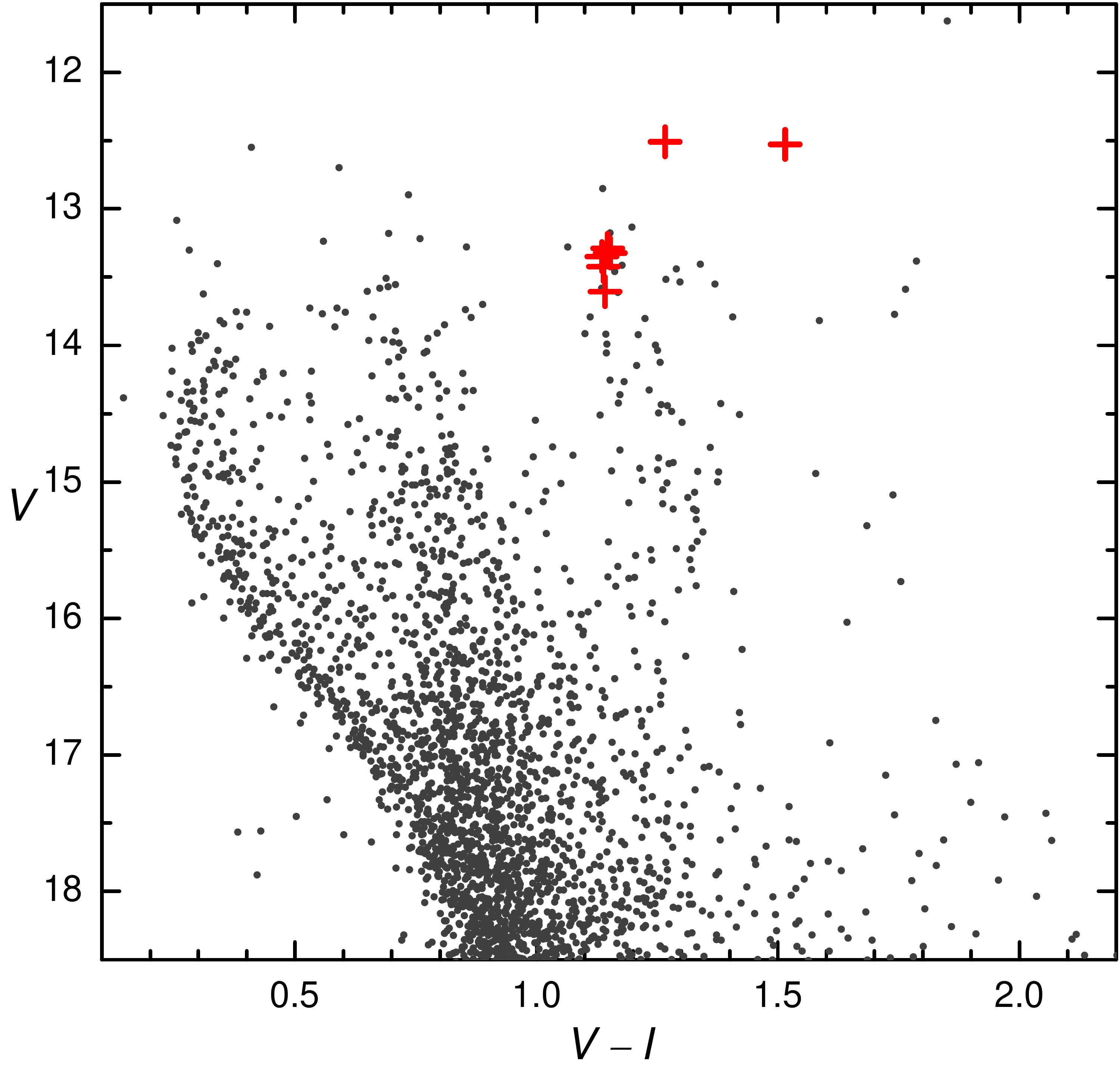}
\caption{Colour-magnitude diagram of the open cluster NGC\,2324. The stars investigated in this work are indicated by crosses. The diagram is based on $VIc$\ photometry by \citet{Kyeong01}.} 
\label{Fig1}
\end{figure}

\begin{figure}
 \includegraphics[width=0.47\textwidth]{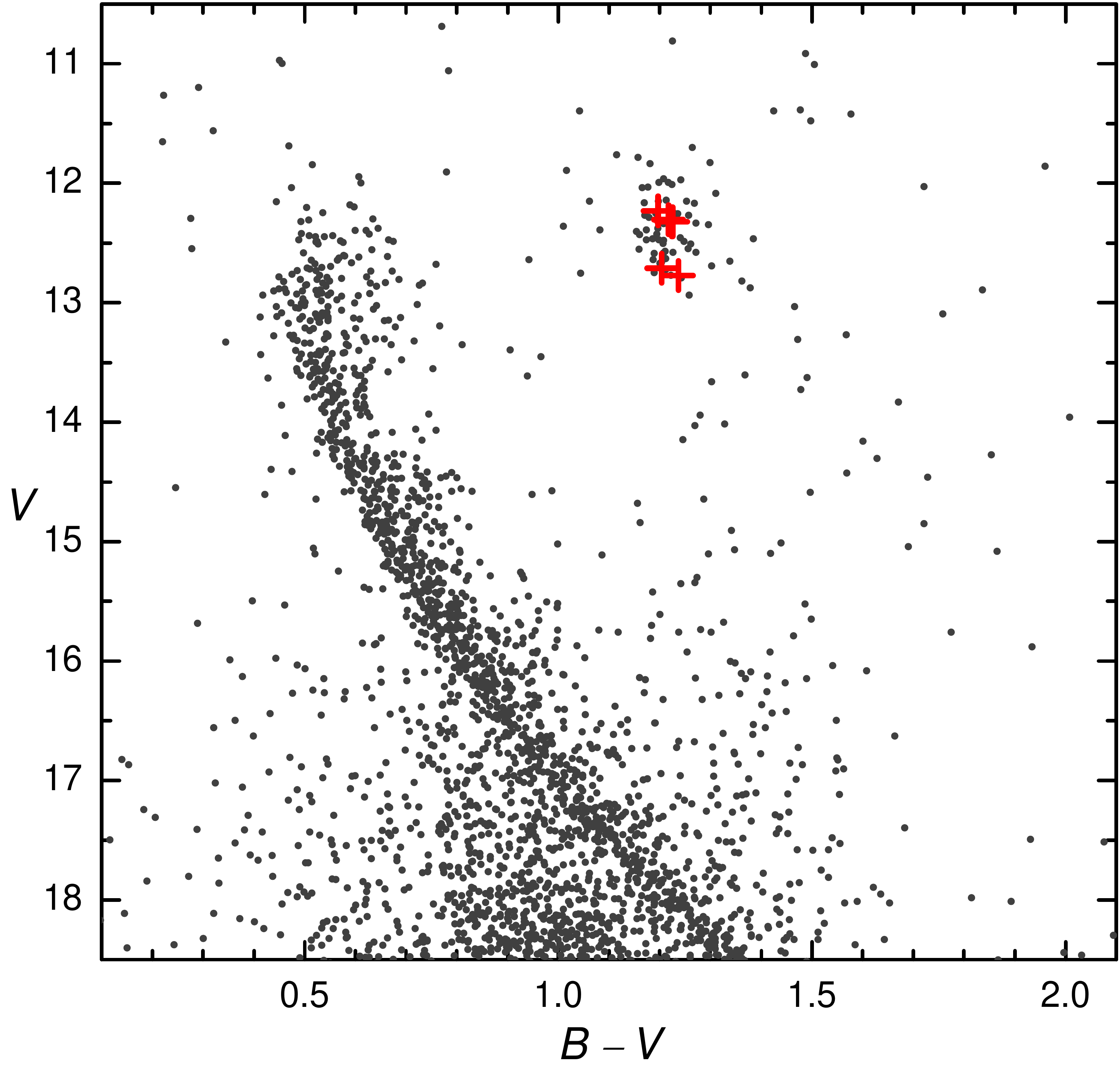}
 \caption{Colour-magnitude diagram of the open cluster NGC\,2447. The stars investigated in this work are indicated by crosses. The diagram is based on $UBV$\ photometry by \citet{Kassis97}.} 
\label{Fig2}
\end{figure}

\begin{figure}
 \includegraphics[width=0.47\textwidth]{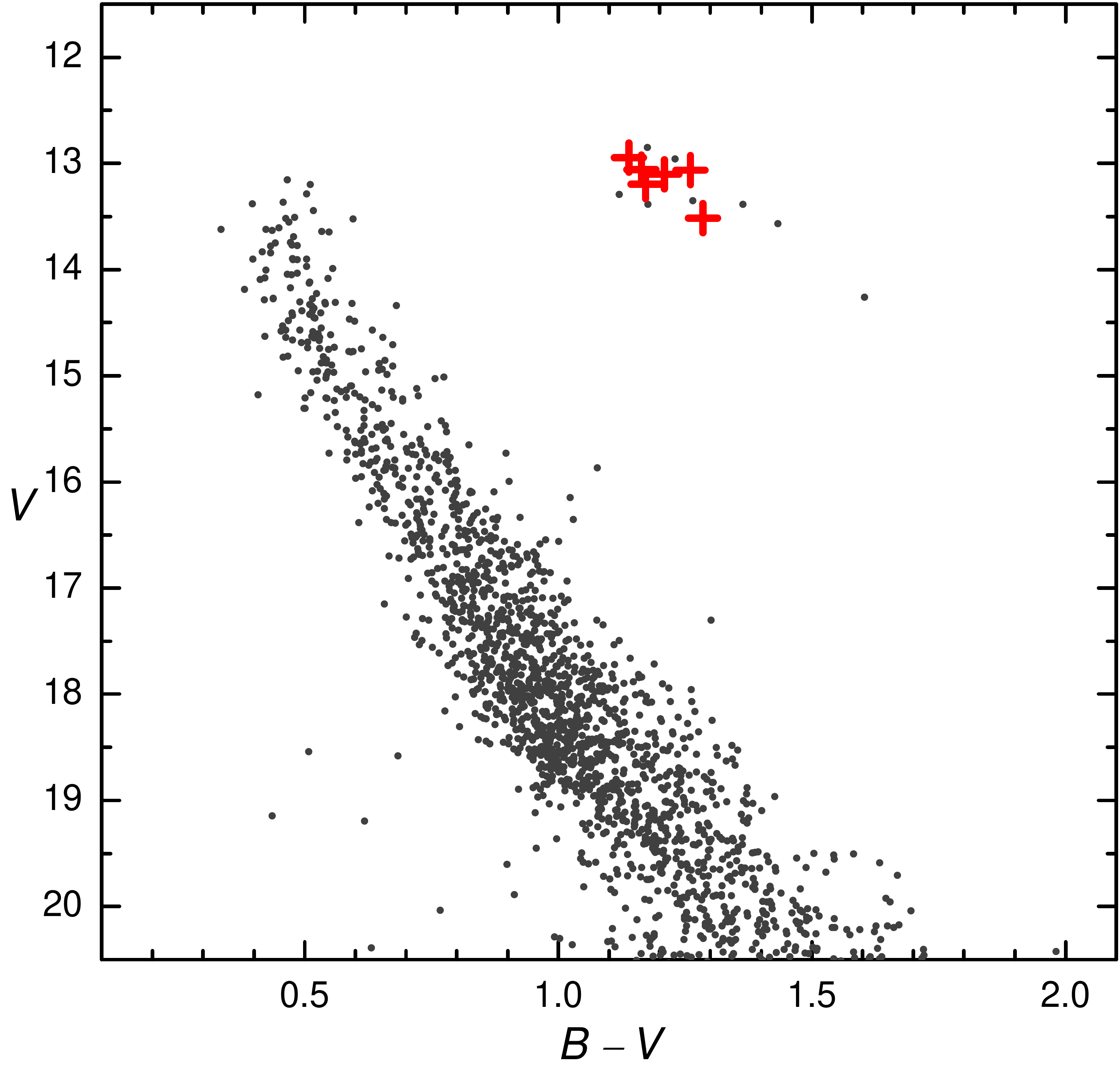}
 \caption{Colour-magnitude diagram of the open cluster NGC\,3960. The stars investigated in this work are 
indicated by crosses. The diagram is based on $BV$\ photometry by \citet{Bragaglia06}.} 
\label{Fig3}
\end{figure}

Photometry of NGC\,2477 was first made by \citet{Eggen61}. \citet{Hartwick72} suggested based on further photometric observations that the metallicity of NGC\,2477 is approximately 1.5 times that of the Hyades and the age is about 1.5~Gyr. \citet{Kassis97} used deep $UBVI$ photometry and isochrone fitting (with ${\rm [Fe/H]}= -0.05\pm0.11$ from \citet{Friel93} and reddening 0.2--0.4 from \citet{Hartwick72} to derive an age of about 1~Gyr and $R_{\rm gc}=8.94$~kpc. \citet{Kharchenko13} reported an approximate distance from the Sun  of 1.5~kpc for this cluster and an age 
equal to 820~Myr. The low-resolution spectroscopy performed by \citet{Friel02} gave ${\rm [Fe/H]}= -0.13\pm 0.10$. The high-resolution spectroscopic metallicity value is ${\rm [Fe/H]}=0.07\pm 0.03$ (\citealt{Bragaglia08}).

NGC\,3960 was identified by \citet{Bergh75}. Its first colour-magnitude diagram was built by \citet{Janes81}. \citet{Prisinzano04} investigated this cluster using $BVI$ filters and determined its age as between 0.9 and 1.4~Gyr. \citet{Bragaglia06} presented $UBVI$ photometry and derived an age of 0.6 to 0.9~Gyr from a comparison of observed and synthetic CMDs, using tracks without and with overshooting, respectively. The higher value is to be preferred; the corresponding distance modulus and reddening place the cluster at $R_{\rm gc}$=7.4~kpc and z=220~pc above the Galactic plane. The metallicity of the cluster has been studied extensively. The photometric metallicity determinations range from ${\rm [Fe/H]}= -0.68$ (determined by \citealt{Geisler92} using Washington photometry) to $-0.06$~dex by \citealt{Piatti98}, who used $DDO$ photometry. Based on low-resolution spectra, \citet{Friel93} determined ${\rm [Fe/H]}= -0.34$, and \citet{Twarog97}) determined ${\rm [Fe/H]}= -0.17$. A high-resolution spectral analysis of three stars by \citet{Bragaglia06} resulted
in  ${\rm [Fe/H]}= -0.12\pm0.04$, and \citet{Sestito06} derived the mean value as ${\rm [Fe/H]}= +0.02\pm0.04$ from the higher signal-to-noise ratio spectra of six stars.

So far, NGC\,2324, NGC\,2477, and  NGC\,3960 had no stars with carbon, nitrogen and oxygen, and carbon isotopic 
$^{12}{\rm C}/^{13}{\rm C}$ ratios determined.

 \begin{table*}
     \centering
\begin{minipage}{190mm}
\caption{Main parameters of the target stars.}
\label{table:1}
\begin{tabular}{lcccccccccc}
\hline\hline
\noalign{\smallskip}
Cluster &       Star    &       $T_{\rm eff}$   &       log $g$ &       $v_{\rm t}$     &       [Fe/H]  &       $v$\,sin \,$i$  &$V$    &       $B-V$   &       $V-I$&  RV \\
        &       ID      &       K       &               &       km s$^{-1}$        &               &       km s$^{-1}$     &mag    &       mag     &       mag     & km s$^{-1}$     \\
\hline                                                                                                                                  
\noalign{\smallskip}                                                                                                                                    
NGC 2324        &       850     &       5100    &       2.20    &       1.21    &       $-0.17$ &       7.5     &13.351 &               &       1.135& 39.07           \\
NGC 2324        &       2603    &       4300    &       1.00    &       1.37    &       $-0.14$ &       3       &12.530 &               &       1.479& 42.14           \\
NGC 2324        &       2225    &       5060    &       2.10    &       1.23    &       $-0.17$ &       6       &13.422 &               &       1.139& 45.85           \\
NGC 2324        &       1006    &       5040    &       2.25    &       1.21    &       $-0.17$ &       3       &13.291 &               &       1.147& 34.91           \\
NGC 2324        &       1992    &       4750    &       1.65    &       1.29    &       $-0.10$ &       4       &12.580 &               &       1.266& 41.78           \\
NGC 2324        &       1788    &       5000    &       2.10    &       1.22    &       $-0.27$ &       7       &13.608 &               &       1.142& 41.82                   \\
NGC 2324        &       2027    &       5000    &       1.84    &       1.26    &       $-0.17$ &       4       &13.325 &               &       1.153& 37.65           \\
\noalign{\smallskip}                                                                                                                                    
\hline                                                                                                                                  
\noalign{\smallskip}                                                                                                                                    
NGC 2477        &       13385&  4980    &       2.80    &       1.14    &       0.05    &               &12.771 &       1.237   &               & 7.02            \\
NGC 2477        &       4221    &       4970    &       2.68    &       1.15    &       0.05    &               &12.231 &       1.197   &               & 8.27    \\
NGC 2477        &       5035    &       5000    &       2.70    &       1.15    &       0.10    &               &12.306 &       1.217   &               & 6.71    \\
NGC 2477        &       3206    &       4950    &       2.66    &       1.16    &       0.05    &               &12.321 &       1.223   &               & 6.80    \\
NGC 2477        &       2061    &       5030    &       2.67    &       1.15    &       0.07    &       1.5     &12.710 &       1.204   &               & 8.06    \\
NGC 2477        &       8039    &       4970    &       2.65    &       1.16    &       0.12    &       1       &12.320 &       1.226   &               & 8.18    \\
\noalign{\smallskip}                                                                                                                                    
\hline                                                                                                                                  
\noalign{\smallskip}                                                                                                                                    
NGC 3960        &       310755 &        4950    &       2.35    &       1.19    &       0.00    &               &13.512 &       1.285   &               & $-24.16$        \\
NGC 3960        &       310756 &        5050    &       2.54    &       1.17    &       0.07    &       1.5     &13.194 &       1.172   &               & $-22.39$        \\
NGC 3960        &       310757 &        4870    &       2.16    &       1.22    &       0.00    &               &13.062 &       1.261   &               & $-21.94$        \\
NGC 3960        &       310758 &        4950    &       2.40 &       1.19    &       0.02    &       2.5     &12.945 &       1.139   &               & $-21.86$        \\
NGC 3960        &       310760 &        5040    &       2.57    &       1.18    &       0.00    &       2.0     &13.060 &       1.164   &               & $-32.87$        \\
NGC 3960        &       310761 &        5000    &       2.45    &       1.18    &       0.02    &       1.5     &13.100 &       1.209   &               & $-22.58$        \\
\hline                                                                                                                                  
\end{tabular}
\end{minipage}
\tablefoot{Star IDs are taken from \citet{Piatti04} for NGC\,2324, from WEBDA for NGC\,2477, and from \citet{Prisinzano04} for NGC\,3960. The atmospheric parameters and radial velocities are from \citet{Bragaglia08} for NGC 2324 and NGC 2477, and from \citet{Sestito06} for NGC 3960 (also adopted by \citealt{Bragaglia08}). $v$\,sin\,$i$ values were derived in this work.}

\end{table*}


   \begin{table}
    \centering
       \caption{Effects on derived abundances, $\Delta$[A/H], resulting from model changes
for the star NGC\,3960\,310757. }
         \label{table:2}
       \[
          \begin{tabular}{lccccc}
             \hline
             \noalign{\smallskip}
Species & ${ \Delta T_{\rm eff} }\atop{ \pm100~{\rm~K} }$ & ${ \Delta \log g }\atop{ \pm0.3 }$ & ${ \Delta v_{\rm t} }\atop{ \pm0.3~{\rm km~s}^{-1}}$ &
             ${ \Delta {\rm [Fe/H]} }\atop{ \pm0.1}$ & Total \\
             \noalign{\smallskip}
             \hline
             \noalign{\smallskip}
C       &       0.01    &       0.03    &       0.00    &            0.04    &       0.03    \\
N       &       0.06    &       0.07    &       0.01    &            0.10    &       0.08    \\
O       &       0.01    &       0.15    &       0.00    &            0.04    &       0.09    \\
C/N     &       0.11    &       0.03&   0.02    &       0.12         &       0.10\\
$^{12}{\rm C}/^{13}{\rm C}$ &   2               &       2               &         0               &       3       &    2    \\
             \hline
          \end{tabular}
       \]
    \end{table}

   \begin{table}
    \centering
       \caption{Effects of derived abundances and isotopic ratios for the star NGC\,3960\,310757, resulting from 
abundance changes of C, N, or O.}
         \label{table:3}
       \[
          \begin{tabular}{lccc}
             \hline
Species & $\Delta$ C & $\Delta$ N & $\Delta$ O \\
 & $\pm0.1$~dex & $\pm0.1$~dex &$\pm0.1$~dex \\
             \hline
             \noalign{\smallskip}
$\Delta$ C                      &       --              &        0.01    &        0.03   \\
$\Delta$ N                      &       0.12    &       --              &        0.06    \\
$\Delta$ O                      &       0.01    &        0.01   &       --              \\
$\Delta$ C/N                    &        0.19   &        0.16   &        0.02    \\
$\Delta ^{12}{\rm C}/^{13}{\rm C}$      &        3              &        3       &        0      \\
             \hline
          \end{tabular}
       \]
    \end{table}

\section{Observations and method of analysis}

Observations were carried out with the multi-object instrument FLAMES (Fiber Large Array Multi-Element Spectrograph) on the Very Large Telescope at the European Southern Observatory, Chile (\citealt{Pasquini02}). The fiber link to UVES (Ultraviolet and Visual \'{E}chelle Spectrograph, \citealt{Dekker00}) was used to obtain high-resolution spectra ($R = 47\,000$). Signal-to-noise ratios of the spectra were between 80 and 190, depending on stellar brightness. Details of observations and reductions are presented by \citet{Bragaglia08} for NGC 2324 and NGC 2477, and by \citet{Sestito06} for NGC 3960. For the analysis we used spectra of seven red clump stars of NGC\,2324, and six stars of NGC\,2477 and NGC\,3960. Figures~\ref{Fig1} -- \ref{Fig3} show the colour-magnitude diagrams of the stars analysed in these clusters.

We analysed the spectra using the same differential analysis technique as in \citet{Drazdauskas16}. All calculations were differential with respect to the Sun. Solar element abundance values were taken from \citet{Grevesse02}. The main atmospheric parameters of the target stars were taken from \citet{Bragaglia08} for NGC 2324 and NGC 2477, and from \citet{Sestito06} for NGC 3960 (also adopted by \citealt{Bragaglia08}), which were determined spectroscopically from the same spectra. For convenience, we present these atmospheric parameters together with photometric data in Table~\ref{table:1}. The star NGC\,2324\,1788 has a slightly lower metallicity than the remaining investigated stars, 
which my be caused by measurement errors (this star is the faintest of the sample). The radial velocity of this star is very 
similar to those of other stars in NGC\,2324.

\begin{figure}
 \includegraphics[width=0.47\textwidth]{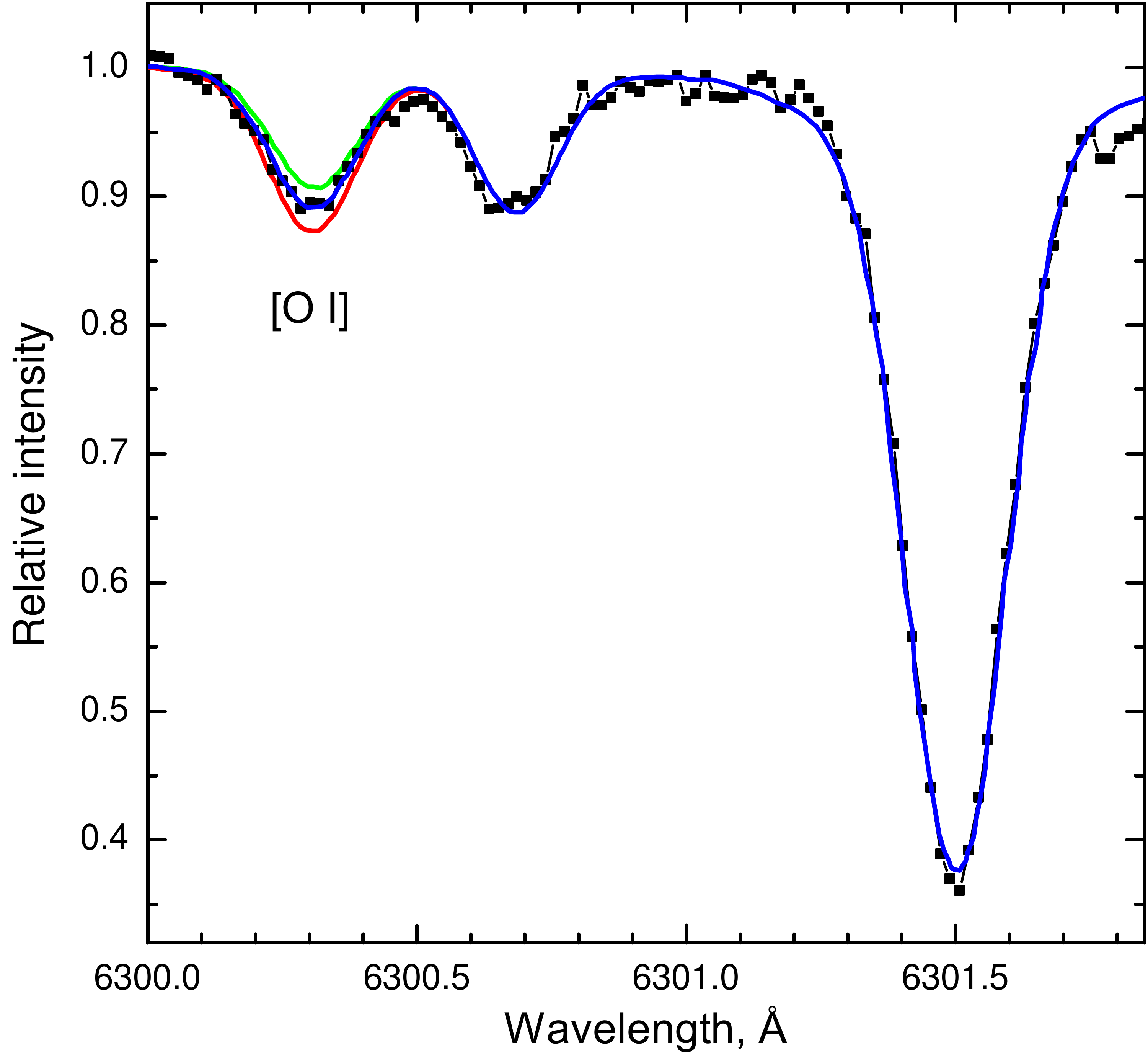}
 \caption{Fit to the forbidden [O\,{\sc i}] line at 6300.3 {\AA} in the spectrum of NGC\,2477\,4221. The observed spectrum is shown as a black line with dots. The synthetic spectra with ${\rm [O/Fe]}=-0.08 \pm 0.1$ are shown as coloured lines.} 
\label{Fig4}
\end{figure}

\begin{figure}
 \includegraphics[width=0.47\textwidth]{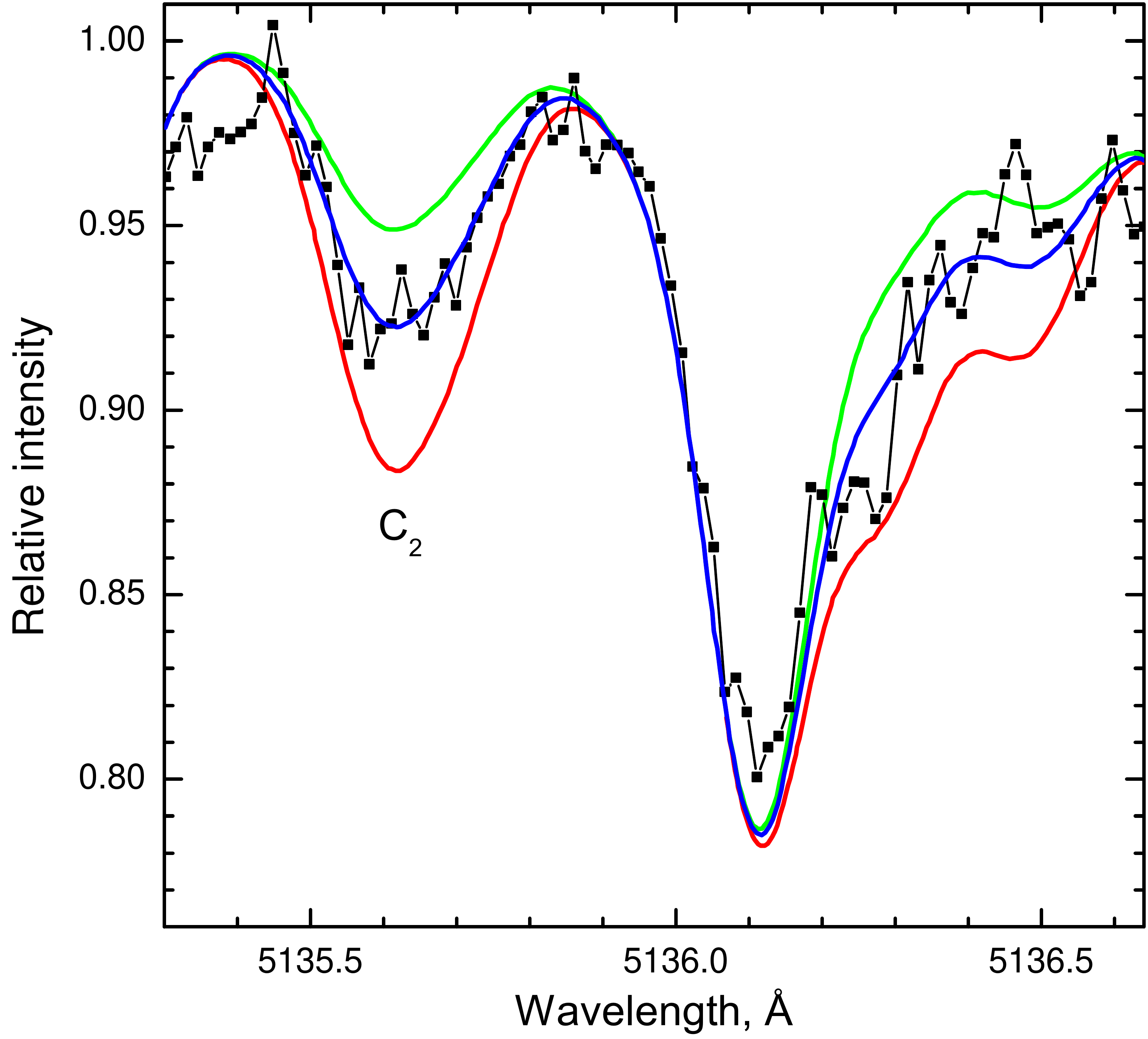}
 \caption{Fit to the ${\rm C}_2$ Swan (1,0) band head at 5135~{\AA} in NGC\,2324\,2027. The observed spectrum is shown as a black line with dots. The synthetic spectra with ${\rm [C/Fe]}=-0.36 \pm 0.1$ are shown as coloured lines.} 
\label{Fig5}
\end{figure}

\begin{figure}
 \includegraphics[width=0.47\textwidth]{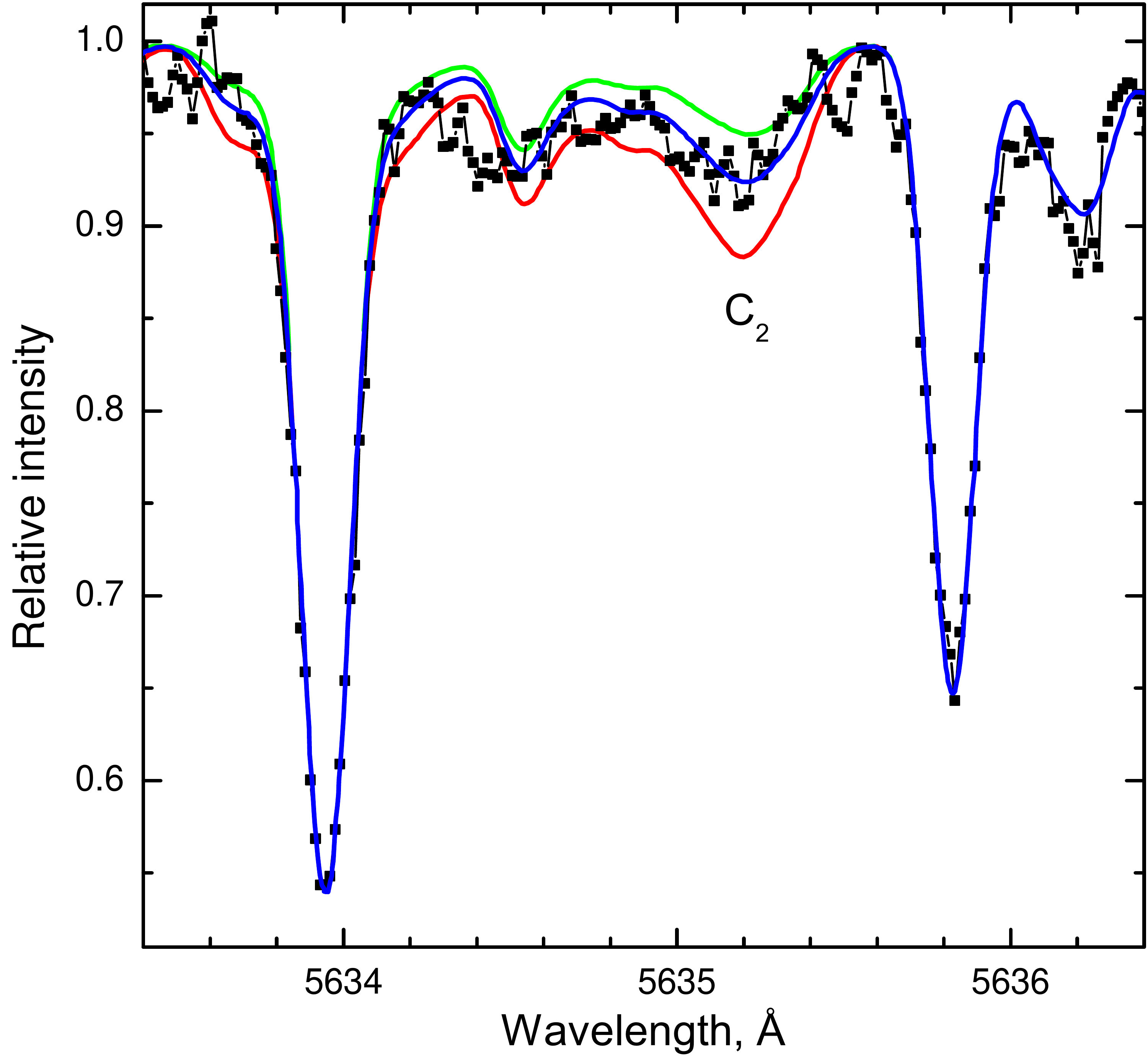}
 \caption{Fit to the ${\rm C}_2$ Swan (0,1) band head at 5635.5~{\AA} in NGC\,2477\,4221. The observed spectrum is shown as a black line with dots. The synthetic spectra with ${\rm [C/Fe]}=-0.26 \pm 0.1$ are shown as coloured lines.} 
\label{Fig6}
\end{figure}
\begin{figure}
 \includegraphics[width=0.47\textwidth]{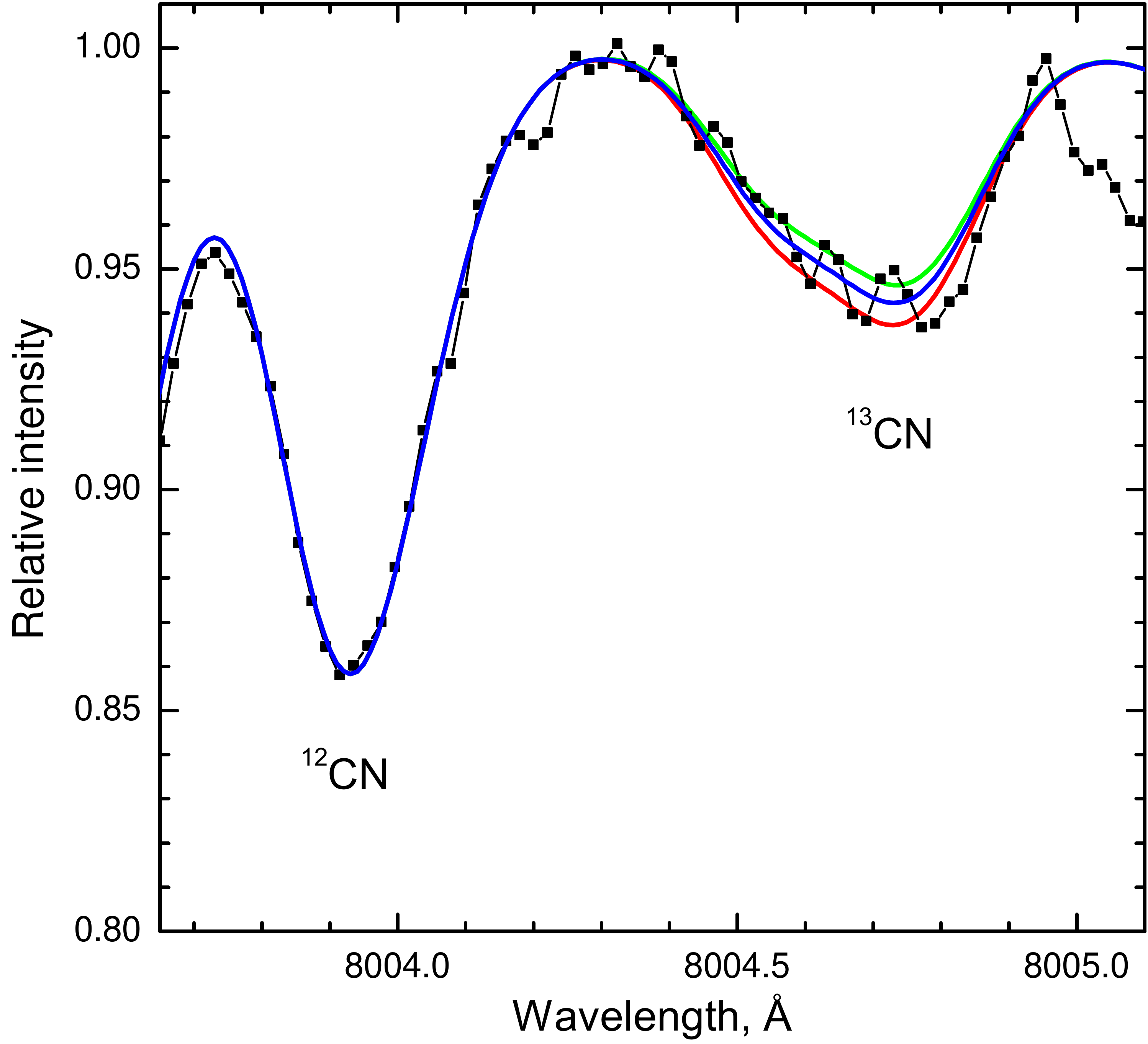}
 \caption{Fit to the CN bands at 8002--8006~\AA{} in NGC\,3960\,310757. The blue line represents [N/Fe]=0.31 and $^{12}{\rm C}/^{13}{\rm C}$=10, the other two lines show $+1$ (green line) and $-1$ (red  line) to the $^{12}{\rm C}/^{13}{\rm C}$ ratio.} 
\label{Fig7}
\end{figure}

Spectral synthesis was used for all abundance determinations and to calculate the $^{12}{\rm C}/^{13}{\rm C}$\ ratio. The program BSYN, developed at the Uppsala University, was used for the spectral syntheses. A set of plane-parallel one-dimensional hydrostatic LTE (local thermal equilibrium) model atmospheres with constant
flux were taken from the MARCS stellar model atmosphere and flux library\footnote{http://marcs.astro.uu.se/} (\citealt{Gustafsson08}).

The Vienna Atomic Line Data Base (VALD, \citealt{Kupka00}) was used to prepare input data for the calculations. Atomic oscillator strengths for the main spectral lines analysed in this study were taken from an inverse solar spectrum analysis (\citealt{Gurtovenko89}).

For the carbon abundance determination in all stars we used two regions: the ${\rm C}_2$ Swan (0,1) band heads at 5135.5~{\AA} and 5635.2~{\AA}. We used the same molecular data of ${\rm C}_2$  as \citet{Gonzalez98}. The oxygen abundance was derived from synthesis of the forbidden [O\,{\sc i}] line at 6300~{\AA}. The $gf$\ values for $^{58}{\rm Ni}$ and $^{60}{\rm Ni}$ isotopic line components, which are blended with the oxygen line, were taken from \citet{Johansson03}. The interval $7980-8010$~{\AA,
which} contains strong CN features, was used to determine the nitrogen abundance and the $^{12}{\rm C}/^{13}{\rm C}$\ ratios. The $^{12}{\rm C}/^{13}{\rm C}$\ ratio was obtained from the $^{13}{\rm C}/^{12}{\rm N}$ feature at 8004.7~{\AA}. The CN molecular data for this wavelength interval were provided by Bertrand Plez.

All the synthetic spectra were calibrated to the solar spectrum by \citet{Kurucz05} to make the analysis differential to the Sun. An instrumental profile was adjusted using the solar spectrum observed on the same instrument as the programme stars. Line broadening by stellar rotation was taken into account for stars when necessary. The highest $v$\,sin\,$i$ values (from 3 to 7.5~km~s$^{-1}$) were found for stars in the youngest open cluster in our sample, NGC\,2324.  We estimated the approximate rotational velocity by looking at stronger surrounding lines in spectral regions around the investigated C N O features. Our roughly estimated upper values of $v$\,sin\,$i$ are presented together with the atmospheric parameters in Table~\ref{table:1}. They may include some influence of macroturbulence.

\begin{table*}
     \centering
\begin{minipage}{190mm}
\caption{Determined abundances and isotopic ratios of the target stars.}
\label{table:4}
\begin{tabular}{lccccccc}
\hline\hline
Star    &       [C/H]   &       $\sigma$ [C/H]  &       [N/H]   &       $\sigma$ [N/H]   &       [O/H]   &       C/N     &       $^{12}$C/$^{13}C$       \\      
\hline                                                                                                                          
\noalign{\smallskip}                                                                                                                            
&&&& NGC\,2324 &&& \\                                                                                                                           
850             &       $-0.47$ &       0.02    &       0.16            &       0.06    &       $-0.39$ &       0.93    &               \\      
2603$^\ast$     &       $-0.57$ &       0.06    &       0.08            &       0.08    &       $-0.42$ &       0.89    &       22      \\      
2225            &       $-0.52$ &       0.01    &       0.10            &       0.03    &       $-0.37$ &       0.95    &               \\      
1006                    &       $-0.45$ &       0.02    &       0.16            &       0.04    &       $-0.22$ &       0.98    &       20      \\      
1992$^\ast$ &   $-0.53$ &       0.02    &       0.21            &       0.06    &       $-0.22$ &       0.72    &       20      \\      
1788                    &   $-0.60$     &       0.06    &       $-0.05$ &       0.02    &       $-0.44$ &       1.12    &               \\      
2027                    &       $-0.53$ &       0.04    &       0.14            &       0.06    &       $-0.53$ &       0.85    &       20      \\      
\hline                                                                                                                          
\noalign{\smallskip}
Average &       $-0.52 \pm0.05$&        & 0.11 $\pm0.08$&               & $-0.37 \pm0.10$&0.92 $\pm0.12$&   $21\pm1$              \\      
\hline                                                                                                                          
\noalign{\smallskip}                                                                                                                            
&&&& NGC\,2477 &&& \\                                                                                                                           
13385   &       $-0.20$ &       0.02    &       0.39    &       0.02    &       $-0.09$ &       1.02    &       18      \\      
4221            &       $-0.21$ &       0.01    &       0.44    &       0.01    &       $-0.03$ &       0.89    &       20      \\      
5035            &       $-0.19$ &       0.03    &       0.51    &       0.05    &       $-0.07$ &       0.79    &               \\      
3206            &       $-0.20$ &       0.04    &       0.49    &       0.06    &       ~~0.05  &       0.81    &       20      \\      
2061            &       $-0.19$ &       0.01    &       0.41    &       0.06    &       $-0.09$ &       1.00    &               \\      
8039            &       $-0.12$ &       0.01    &       0.52    &       0.05    &       ~~0.03  &       0.91    &       20      \\      
\hline                                                                                                                          
\noalign{\smallskip}
Average &       $-0.19$ $\pm$ 0.03      &       &       0.46 $\pm$ 0.05      &       &       $-0.03$ $\pm$ 0.06      &       0.91 $\pm$ 0.09& $20\pm1$    \\      
\hline                                                                                                                          
\noalign{\smallskip}                                                                                                                            
 &&&& NGC\,3960 &&& \\                                                                                                                          
 310755  &      $-0.37$ &       0.02    &       0.23    &       0.05    &       $-0.13$ &       1.00    &               \\      
310756   &      $-0.33$ &       0.02    &       0.44    &       0.03    &       $-0.16$ &       0.68    &       12      \\      
310757   &      $-0.47$ &       0.02    &       0.31    &       0.05    &       $-0.25$ &       0.66    &       10      \\      
310758   &      $-0.35$ &       0.03    &       0.33    &       0.03    &       $-0.17$ &       0.83    &       20      \\      
310760   &      $-0.38$ &       0.01    &       0.35    &       0.02    &       $-0.25$ &       0.74    &       16      \\      
310761   &      $-0.32$ &       0.04    &       0.34&   0.04    &       $-0.08$ &       0.87    &       18      \\      
\hline                                                                                                                          
\noalign{\smallskip}
Average &       $-0.37$ $\pm$ 0.05      &       &       0.33 $\pm$ 0.07      &       &       $-0.17$ $\pm$ 0.07      &       0.80 $\pm$ 0.13      & $16\pm4$      \\
\hline                                                                                                                                                                          
\end{tabular}
\end{minipage}
\tablefoot{Carbon abundances were determined from two lines. Nitrogen abundances from 3 to 8 lines, and oxygen 
 abundance from a single line.} 
\end{table*}

Fitting of individual lines depends on several factors, including uncertainties on atomic parameters, continuum placement variations, and the fitting of synthetic spectra to each line. Secondly, there are errors that affect all measured lines simultaneously, such as uncertainties in the stellar atmospheric parameters. All stars are quite similar in their parameters and the quality of their spectra, therefore we chose one star (NGC\,3960\,310757) and calculated the effect of the assumed uncertainties of the atmospheric parameters on the abundance estimates (Table~\ref{table:2}). Considering the given deviations from the parameters, we see that the abundances are not affected strongly. Since the CNO
abundances are also bound together by molecular equilibrium in the stellar atmospheres, we also investigated the effect of an error in one of them on the abundance determination of another (Table~\ref{table:3}). 

\section{Results and discussion}

The abundances of carbon, nitrogen, and oxygen relative to hydrogen [El/H] (we use here the customary spectroscopic notation [X/Y]$\equiv \log_{10}(N_{\rm X}/N_{\rm Y})_{\rm star} -\log_{10}(N_{\rm X}/N_{\rm Y})_\odot$)  and $\sigma$ (the line-to-line scatter) are listed
along with the C/N and $^{12}$C/$^{13}$C number ratios in Table~\ref{table:4}.

Carbon and nitrogen abundances together with C/N and $^{12}{\rm C}/^{13}{\rm C}$ ratios have been analysed in red giant stars of open clusters, globular clusters, Galactic halo and discs as indicators of mixing processes for decades (e.g. \citealt{Dearborn75, Tomkin76, Lambert77, Lambert81, Luck78,  Suntzeff81, Sneden86, Gilroy89, Gratton00, Tautvaisiene01, Tautvaisiene10, Tautvaisiene13, Shetrone03, Spite06, Smiljanic09,  Mikolaitis10, Angelou12, Masseron15, Drazdauskas16}, and references therein).

\begin{figure}
 \includegraphics[width=0.47\textwidth]{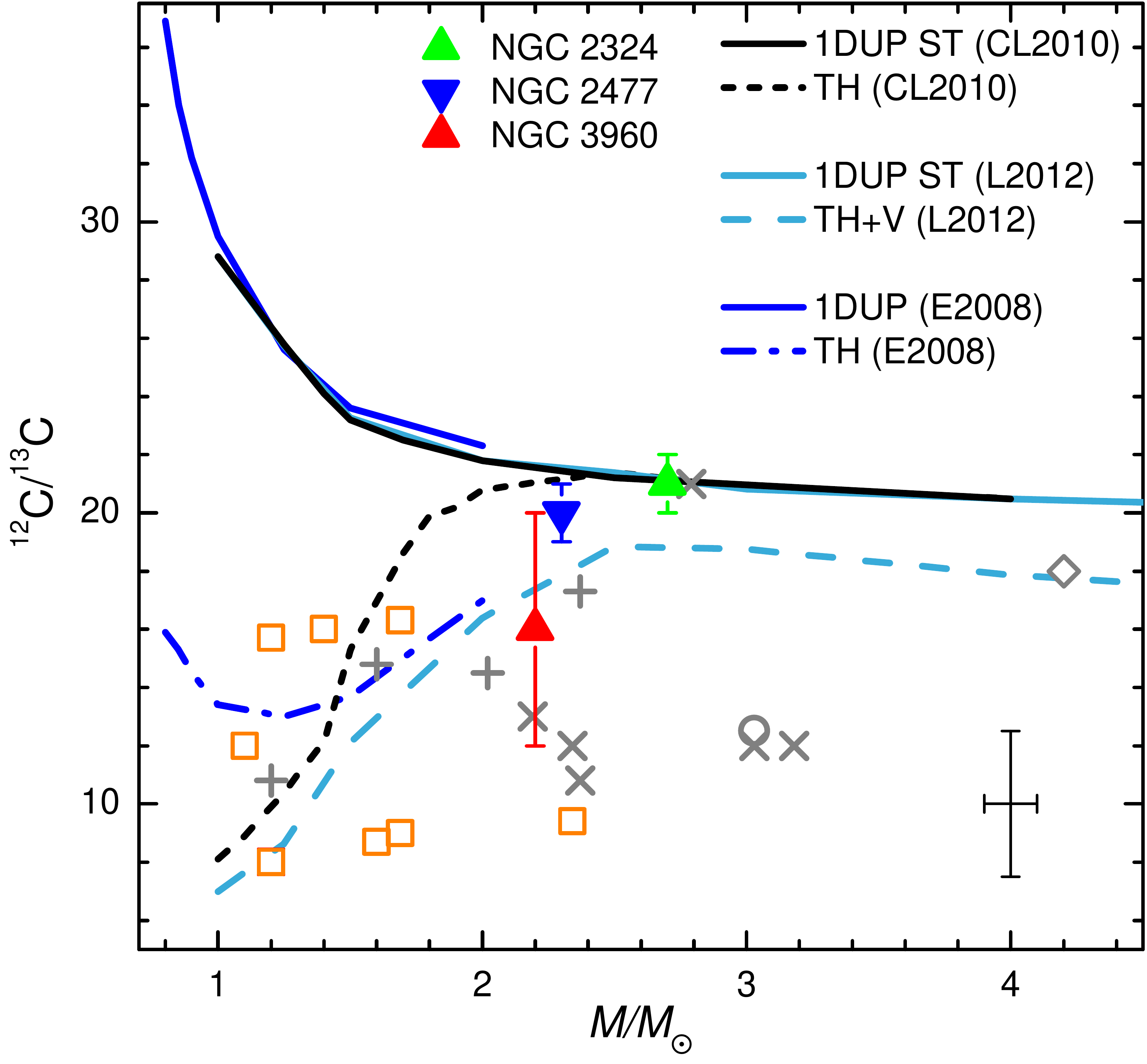}
 \caption{Average $^{12}$C/$^{13}$C ratios in clump stars of open clusters as a function of stellar turn-off mass. Our work is represented by coloured triangles and bars indicating a star-to-star scatter. Open squares are for determinations by \citet{Drazdauskas16}, \citet{Mikolaitis10, Mikolaitis11a, Mikolaitis11b, Mikolaitis12}, and \citet{Tautvaisiene00, Tautvaisiene05}, crosses for \citet{Smiljanic09}, a diamond for \citet{Santrich13}, a circle is for \citet{Luck94}, and plus signs for \citet{Gilroy89}. The solid lines represent standard dredge-up models \citep{Eggleton08, Charbonnel10, Lagarde12}), the dash-dotted line represents the thermohaline mixing model by \citet{Eggleton08}, the short-dashed line is for a model including thermohaline mixing \citet{Charbonnel10}, and the long-dashed line represents a model that includes both the thermohaline and rotational effects (\citealt{Lagarde12}). A typical error bar is indicated (\citealt{Smiljanic09, Gilroy89}).} 
 \label{Fig8}
\end{figure}

\begin{figure}
 \includegraphics[width=0.47\textwidth]{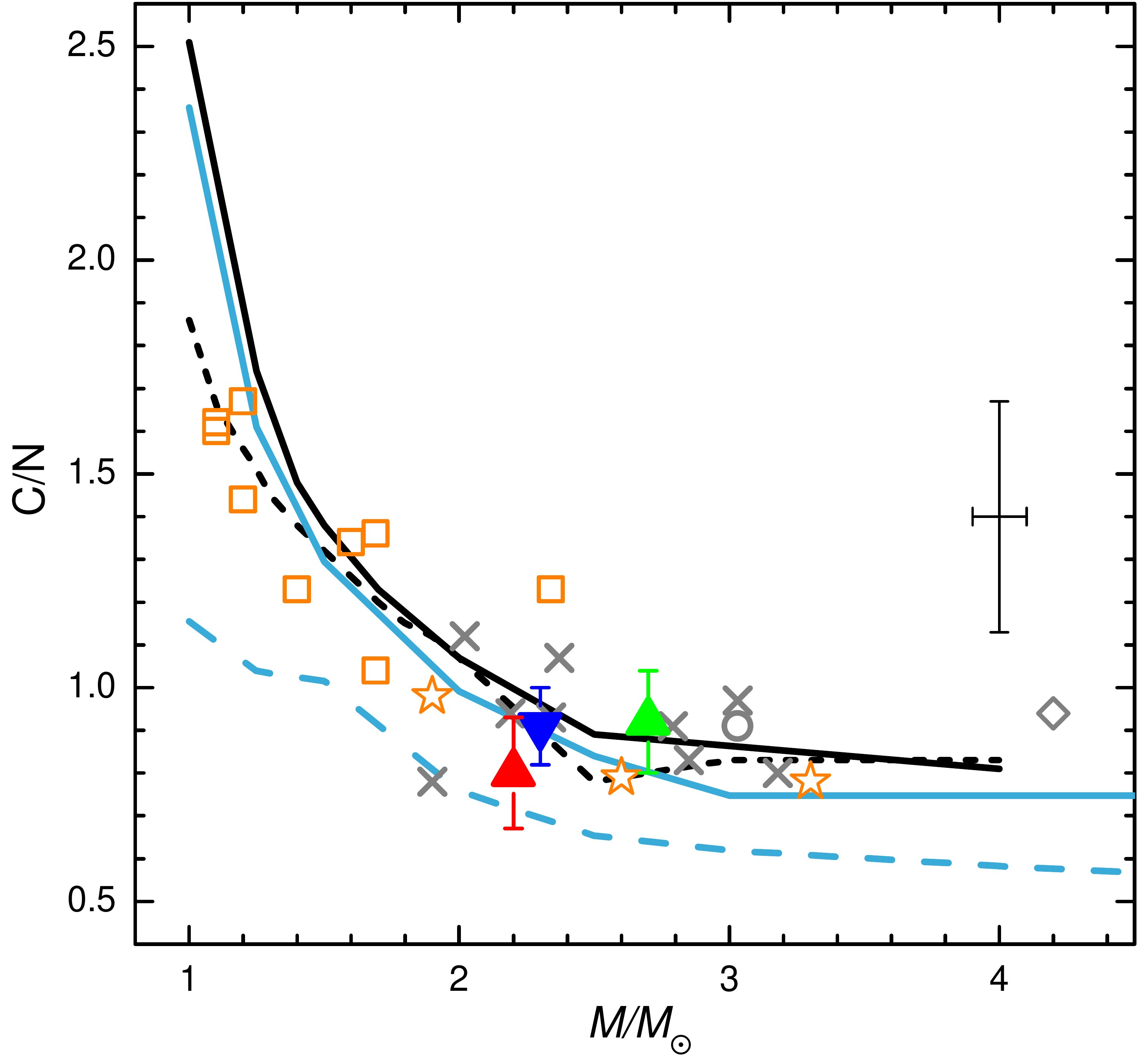}
 \caption{Average C/N ratios in clump stars of open clusters a a function of stellar turn-off mass. The meaning of the symbols is the same as in Fig.~\ref{Fig8}. Results from \citet{Tautvaisiene15} are shown as open stars.} 
\label{Fig9}
 \end{figure}

In Figs.~\ref{Fig8}--\ref{Fig9} we compare our results with the
modelled values of the $^{12}{\rm C}/^{13}{\rm C}$ and C/N ratios in respect of stellar turn-off masses.  
The mean metallicities of the clusters (Table~1) and the ages from the literature were used to compute the PARSEC isochrones \citep{Bressan12} and to determine turn-off masses for the target clusters. We derived a turn-off mass of $\sim 2.7~M_{\odot}$ with the adopted age of 440~Myr (determined by \citealt{Piatti04}) for NGC\,2324, $\sim 2.3~M_{\odot}$ with the age of 820~Myr (\citealt{Kharchenko13}) for NGC\,2477, and $\sim 2.2~M_{\odot}$ with the age of 900~Myr (\citealt{Bragaglia06}) for NGC\,3960. 

We also included in the comparison the observational results from other studies of clump stars of open clusters \citep{Gilroy89, Luck94, Tautvaisiene00, Tautvaisiene05, Tautvaisiene15, Mikolaitis10, Mikolaitis11a, Mikolaitis11b, Mikolaitis12, Smiljanic09, Santrich13, Drazdauskas16}. To produce the average values shown in the figures, we used only red clump stars because they provide information on the final composition changes, after all the evolution along the red giant branch. The two stars lying above the clump in NGC\,2324 were included in the average since their results are similar to other investigated stars.

We used the theoretical models by \citet{Eggleton08}, \citet{Charbonnel10}, and \citet{Lagarde12} for the comparison. They provide quantitative values representing the first dredge-up, thermohaline (TH), and thermohaline and rotation (TH+V) induced mixing. \citet{Eggleton08} estimated the mixing speed with their formula for the diffusion coefficient and found that a range of three orders of magnitude in their free parameter can lead to the observed levels of $^{12}{\rm C}/^{13}{\rm C}$. Their predicted value of the $^{12}{\rm C}/^{13}{\rm C}$ ratio for a solar-metallicity 2~$M_{\odot}$ star at the RGB tip is 17. A more recent model of the thermohaline-induced mixing by \citet{Charbonnel10} lists for the same stars a higher value of about 20. The model of \citet{Charbonnel10} of thermohaline instability induced mixing and the model by \citet{Eggleton08} are both based on the ideas of \citet{Eggleton06} and \citet{Ulrich72}. It includes developments by \citet{Charbonnel07}. \citet{Eggleton06} found a mean molecular weight ($\mu$) inversion in their $1~M_{\odot}$ stellar evolution model, which occurred after the so-called luminosity bump on the RGB, when the hydrogen-burning shell reaches the chemically homogeneous part of the envelope. The $\mu$-inversion is produced by the reaction $^3{\rm He(}^3{\rm He}, 2p)^4{\rm He}$, as predicted in \citet{Ulrich72}. It does not occur earlier because the magnitude of the $\mu$-inversion is low and negligible compared to a stabilising $\mu$-stratification. \citet{Charbonnel07} computed stellar evolution models including the ideas of \citet{Kippenhahn80}, who extended Ulrich's equations to the case of a non-perfect gas. \citet{Charbonnel07} also introduced a double diffusive instability (i.e. thermohaline convection) and showed its importance in the chemical evolution of red giants. This mixing connects the convective envelope with the external wing of the hydrogen-burning shell and induces surface abundance modifications in evolved stars (\citealt {Charbonnel10}).

The mean values of $^{12}{\rm C}/^{13}{\rm C}$ in NGC\,2324, NGC\,2477, and  NGC\,3960 are $21\pm1$, $20\pm1$, and $16\pm4$, respectively. 
Thus we see that the mean values of carbon isotope ratios of NGC\,2324 and NGC\,2477 agree well with the model of \citet {Charbonnel10}  of pure thermohaline-induced mixing. NGC\,3960 has the lower mean carbon isotope ratio, which is most probably also affected by rotation-induced mixing. The model that includes both thermohaline- and rotation-induced mixing (\citealt{Lagarde12}) also gives $^{12}{\rm C}/^{13}{\rm C}=16$. The rotation velocity in the model corresponds to 30\% of the critical velocity at the zero-age main sequence (ZAMS, Lagarde et al. 2014, the rotation-induced mixing modifies the internal chemical  structure of main-sequence stars, although its signatures are revealed only later in the stellar evolution). 

\begin{figure*}
 \includegraphics[width=0.90\textwidth]{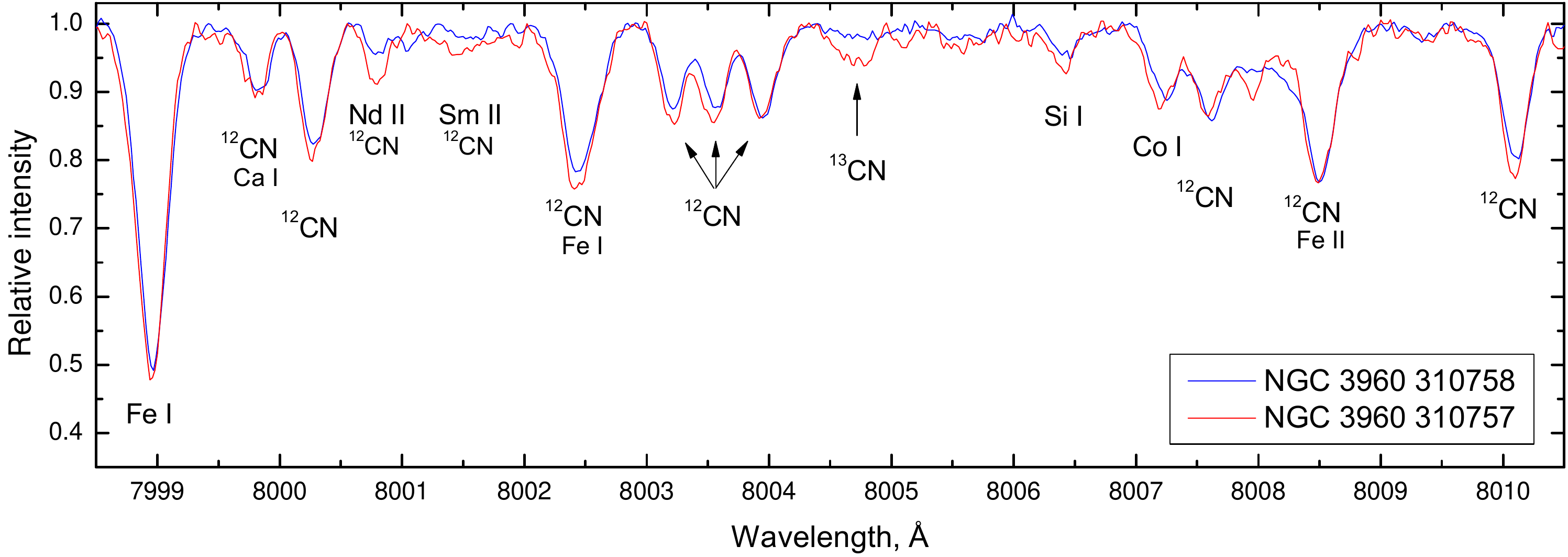}
 \caption{Spectra of NGC\,3960\,310757 (red line), which has a carbon isotope ratio equal to 10,  and spectrum of NGC\,3960\,310758 (blue line), which has a value of 20.
} 
\label{Fig10}
 \end{figure*}

NGC\,3960 is quite interesting. All the investigated stars are at the red clump stage with quite similar atmospheric parameters, but the $^{12}{\rm C}/^{13}{\rm C}$ values span from 10 to 20, and this scatter is natural. In Fig.~\ref{Fig10} we compare a spectrum of NGC\,3960\,310757 with a carbon isotope ratio equal to 10  and a spectrum of NGC\,3960\,310758 with a value of 20. The difference in their $^{13}{\rm CN}$ bands is evident. 

Other open clusters with intermediate turn-off masses have even lower carbon isotope ratios, which is probably due to stronger rotation-induced or other types of mixing (see Fig.~\ref{Fig8}), and the TH+V model is certainly closer to them than the pure thermohaline mixing model. It is unclear, however, why we do not see the same behaviour when comparing their C/N ratios with the corresponding theoretical models. In Fig.~\ref{Fig9} we show the observational results lying close to the 1DUP and thermohaline mixing models, which are very similar at these turn-off masses. The mean C/N values of NGC\,2324 and NGC\,2477 ($0.92\pm0.12$ and $0.91\pm0.20$, respectively) are close to the mentioned models. 

So far, the model including both thermohaline- and rotation-induced mixing (\citealt{Lagarde12}) predicted lower than observed C/N ratios at all turn-off masses of open clusters, but the mean C/N ratio of NGC\,3960, $0.80\pm0.13$, is comparable with this model. In the future more open clusters may be found with C/N values as low as predicted by this model, with a ZAMS rotation velocity of 30\% of the critical velocity or higher.    

\subsection{Oxygen}

While carbon and nitrogen abundances are susceptible to the evolutionary processes that occur inside a star, oxygen, instead, is not. The abundances of oxygen remain more or less constant from the time a star forms and can be used to trace and study the Galactic chemical evolution.

\begin{table}
    \centering
\caption{Galactocentric distances of clusters}
\label{table:5}
\begin{tabular}{lrrccr}
\hline\hline
\noalign{\smallskip}
Cluster    &    \multicolumn{1}{c}{$ l $}    & \multicolumn{1}{c}{$ b $}    &    $d_{\odot}$  & Ref.  & $R_{\rm gc}$\\
        &    \multicolumn{1}{c}{deg}    & \multicolumn{1}{c}{deg}    &    kpc    &  &      kpc \\
\hline
\noalign{\smallskip}
Collinder 261   &    301.684    &    $-05.528$    & 2.75  &   1    & 7.0\\
IC 4651            &    340.088    &    $-07.907$ & 0.9    & 2  &     7.2    \\
Melotte 66       &    259.559    &    $-14.244$    & 4.5   &   2  &   10.0 \\
NGC 2506        &    230.564    &  $~~~09.935$  & 3.3  &   1   &  10.4\\
NGC 2324        &    213.447    &   $~~~03.297$ & 3.8 &   3   & 11.5 \\
NGC 2477        &    253.563    &    $-05.838$    & 1.5  &  2   &   8.6 \\
NGC 3960        &    294.367    &   $~~~06.183$ & 2.1 &  4   & 7.4 \\
NGC 4609        &    301.895    &    $-00.142$    & 1.3  &  2    &   7.5 \\
NGC 4815        &    303.625    &    $-02.097$    & 2.5  & 5  &    6.9 \\
NGC 5316        &    310.229    &  $~~~00.115$  & 1.2  &  2    &  7.4 \\
NGC 6134        &    334.917    &    $-00.198$    & 1.0  & 6    &   7.1 \\
NGC 6253        &    335.460    &    $-06.251$    & 1.6  &  1   &   6.6 \\
NGC 6705        &    27.307    &    $-02.776$      & 1.9  &  7 & 6.3 \\
\hline                                                                                                                                                                           
\end{tabular}
\tablefoot{Distances from the Sun were taken from (1) \citet{BragagliaTosi06},  (2) \citet{Kharchenko13}, (3) \citet{Piatti04}, (4) \citet{Bragaglia06}, (5) \citet{Friel14}, (6) \citet{Ahumada13}, (7) \citet{Cantat14}. The $ R_{\rm gc} $ values were computed with $R _{\rm gc \odot} $= 8.0~kpc.} 
\end{table}

\begin{figure}
 \includegraphics[width=0.47\textwidth]{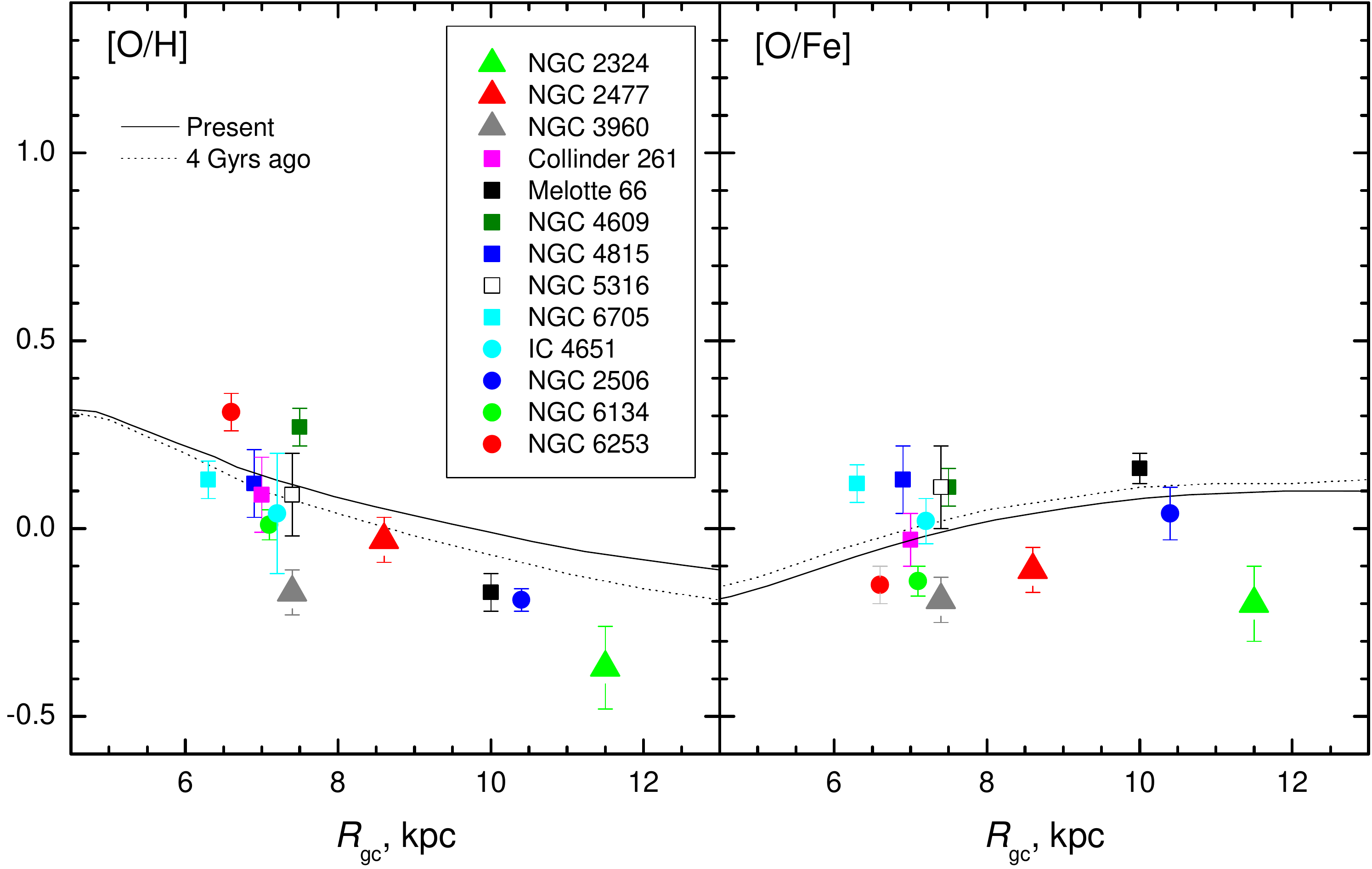}
 \caption{Mean oxygen abundances in relation to the Galactocentric distance compared to the theoretical models by \citet{Magrini09}. The error bars indicate the scatter of stellar abundances inside each cluster. See text for more explanations.} 
\label{Fig11}
 \end{figure}

In Fig.~\ref{Fig11} we plot the determined mean oxygen abundances with respect to Galactocentric distances. We also show the theoretical models by \citet{Magrini09} and include other observational results of open clusters that have recently been investigated by our group 
(\citealt{Mikolaitis10, Mikolaitis11a, Mikolaitis11b, Mikolaitis12, Tautvaisiene15, Drazdauskas16}). For several open clusters we recalculated their Galactocentric distances in a homogeneous way because some of the earlier determinations were based on the solar $R_{\rm gc} = 8.5$~kpc, while we decided to use $R _{\rm gc \odot} $= 8.0~kpc as follows from recent studies (cf. \citealt{Malkin13, Zhu13}).  
Sources of the open cluster distances from the Sun and the resulting Galactocentric distances are presented in Table~\ref{table:5}. Our data points cover $R_{\rm gc}$ from about 6 to 12~kpc. 

There is a clearly visible correlation between the [O/H] abundances and the Galactocentric distance in agreement with previous studies. 
There are more than a few recent studies on the oxygen abundance gradients in our Galaxy related to the Galactocentric distance (see \citealt{Magrini09, Jacobson09, Luck11, Yong12, Korotin14, Martin15, Magrini15} and references therein). All these studies agree that there is a slight decrease in the oxygen abundance with increasing Galactocentric distance. 

Some authors have suggested a bimodal distribution of the oxygen abundances (\citealt{Costa04, Magrini09, Yong12, Korotin14} and references therein) with a flattening starting at around 9--13~kpc from the Galactic centre.   
 \citet{Korotin14}, for example, suggest a double linear distribution with a slope of $-0.056$~dex\,kpc$^{-1}$ until about $R_{\rm gc}=12$~kpc and a flatter one of $-0.033$~dex\,kpc$^{-1}$ at more distant radii. 
Other authors have provided a linear fit for the oxygen abundances without any apparent changes towards the outer regions of the Galaxy. The slopes vary in different studies from $-0.026$~dex\,kpc$^{-1}$ in Cunha et al. (2016) to $-0.055$~dex\,kpc$^{-1}$ in \citet{Luck11} or $-0.06$~dex\,kpc$^{-1}$ in \citet{Rudolph06}.  Our modest sample of open clusters supports studies with the larger slopes, but we cover only about 6~kpc, centred on the position of the Sun.  
  
[O/Fe] trends are also under investigation. We see no apparent correlation of [O/Fe] with $ R_{\rm gc} $ in our sample of clusters, which agrees with other studies. \citet{Jacobson09} analysed the open clusters with $ R_{\rm gc}$ around 10--13~kpc and concluded that any visible trends for the [O/Fe] versus the Galactocentric distance are most likely not real. \citet{Luck11} found a modest increase in [O/Fe] (as well as increasing scatter) towards the outer regions of the Galaxy in Cepheids.

\section{Summary and conclusions}

We have determined CNO abundances and $^{12}$C/$^{13}$C ratios in 19 evolved stars of the open clusters NGC\,2324, NGC\,2477, and  NGC\,3960, which have turn-off masses of about 2~$M_{\odot}$.  

Our results can be summarized as follows: 
\begin{itemize}
\item
The mean values of the $^{12}{\rm C}/^{13}{\rm C}$ and C/N ratios in NGC\,2324 and NGC\,2477 agree well with the first dredge-up and thermohaline-induced extra-mixing models by \citet{Charbonnel10}, which are similar for stars with a turn-off mass of about 2~$M_{\odot}$. 

\item
The $^{12}{\rm C}/^{13}{\rm C}$ ratios in the investigated clup stars of NGC\,3960 span from 10 to 20. The mean carbon isotope and C/N ratio values are close to predictions of the model in which the thermohaline- and rotation-induced (if rotation velocity at ZAMS was 30\% of the critical velocity) extra-mixing act together (\citealt{Lagarde12}).

\item The mean values of [O/H] agree with previous studies in exhibiting decreasing values with increasing Galactocentric distances. [O/Fe] show no apparent correlation with the $ R_{\rm gc} $ in our sample of clusters.
\end{itemize}

Theoretical models of stellar evolution including various mechanisms of material mixing need to be continuously developed. For instance, in the model by \citet{Lagarde12}, the convective envelope was assumed to rotate as a solid body through the evolution, the transport coefficients for chemicals associated with thermohaline- and rotation-induced mixing were simply added in the diffusion equation, and the possible interactions between the two mechanisms were not considered. \citet{Wachlin11} found that to reproduce the observed abundances of red giant branch stars close to the luminosity bump, thermohaline mixing efficiency has to be artificially increased by about four orders of magnitude from what is predicted by recent 3D numerical simulations of thermohaline convection close to astrophysical environments. Numerous shortcomings of various theoretical models were listed by \citet{Canuto11a, Canuto11b, Canuto11c, Canuto11d, Canuto11e}. Like any new paradigm, thermohaline and other types of mixing are stimulating subsequent theoretical and observational studies. 

Nevertheless, carbon and nitrogen are sensitive to evolutionary mixing processes, and an increasing number of open clusters with known 
CNO abundances will be useful for Galactic evolution studies as well. A recent study of Galactic field stars by  \citet{Masseron15} clearly showed this. 

\begin{acknowledgements}
This research has made use of the WEBDA database (operated at the Department of Theoretical Physics and Astrophysics of the Masaryk University, Brno), of SIMBAD (operated at CDS, Strasbourg), of VALD (\citealt{Kupka00}), and of NASA’s Astrophysics Data System. Bertrand Plez (University of Montpellier II) and Guillermo Gonzalez (Washington State University) were particularly generous in providing us with atomic data for CN and C$_2$ molecules, respectively. This work was partly supported (AD, GT, R\v{Z}) by the grant from the Research Council of Lithuania (MIP-082/2015). Partial support was received (AB, SR) from the Italian PRIN MIUR 2010-2011, project "The Chemical and Dynamical Evolution of the Milky Way and Local Group Galaxies".

\end{acknowledgements}

\bibliographystyle{aa} 
\bibliography{References2.bib} 

\end{document}